\documentclass[12pt]{article}%
\usepackage{amsmath,latexsym}
\usepackage{graphicx}
\usepackage{amsmath}
\usepackage{amsfonts}
\usepackage{amssymb}%
\begin{document}

\title{{\Huge On the energy of charged black holes in generalized dilaton-axion
gravity }}
\author{{\large I. Radinschi \thanks{radinschi@yahoo.com}, Farook
Rahaman\thanks{farook\_rahaman@yahoo.com}, and Asish
Ghosh\thanks{farook\_rahaman@yahoo.com}}
\and $^{\dag}$ {\small Department of Mathematics, Jadavpur University, Kolkata -
700032, India}
\and $^{\ast}${\small Department of Physics, "Gh. Asachi" Technical University,}
\and Iasi, 700050, Romania
\and $^{\ddag}${\small Department of Mathematics, Jadavpur University, Kolkata -
700032, India}}
\maketitle
\date{}

\begin{abstract}
In this paper we calculate the energy distribution of some charged black holes
in generalized dilaton-axion gravity. The solutions correspond to charged
black holes arising in a Kalb-Ramond-dilaton background and some existing
non-rotating black hole solutions are recovered in special cases. We focus our
study to asymptotically flat and asymptotically non-flat types of solutions
and resort for this purpose to the M\o ller prescription. Various aspects of
energy are also analyzed.

\end{abstract}

\section{Intoduction}

In the recent years, a wide interest have been focused on numerous efficient
and precise tools, such as superenergy-tensors [1]-[2], energy-momentum
complexes, quasi-local expressions [3] and the tele-parallel theory of
gravitation [4] for the study of energy-momentum localization.

In General Relativity, the problem of localization of energy using
energy-momentum complexes was discussed first by Einstein who constructed his
pseudotensor [5], and other prescriptions were elaborated later by Landau -
Lifshitz [6], Papapetrou [7], Bergmann-Thompson [8], Weinberg [9],
Qadir-Sharif [10] and M\o ller [11]. Among these prescriptions, the M\o ller
definition is the only one which can be applied to any coordinate system,
since the other energy-momentum complexes generate meaningful results only in
the case of the quasi-Cartesian coordinates. Light has been shed upon the
topic of energy-momentum localization in the last two decades and the
pseudotensorial definitions have also been employed for computing the energy
in the case of some 2+1 and 2 dimensional space-times, emphasizing the fact
that different pseudotensorial definitions can yield the same expression for
the energy distribution of a given space-time [12]. We may thus notice that,
in many cases, the energy-momentum complexes produce the same results as their
tele-parallel versions [13]. Virbhadra came up with an important result and
proved that using different energy-complexes (ELLPW) it is possible to obtain
the same result for a general non-static spherically symmetric metric of the
Kerr-Schild class [14]. In addition, these definitions (ELLPW) are compliant
with the quasi-local mass definition given by Penrose [15] and verified by Tod
[16] in the case of a general non-static spherically symmetric metric of the
Kerr-Schild class. Nevertheless, these definitions disagree for the most
general non-static spherically symmetric metric (Virbhadra [14]). We should
also mention the significant results obtained by several authors with the
M\o ller prescription [17]-[18]. Moreover, this definition is considered from
the viewpoint of Lessner [19] as an accurate and powerful tool for energy
localization in General Relativity. Supporting the Lessner opinion and the
meaningful results obtained by several researchers, Chang, Nester and Chen
[20] stressed the fact that the energy-momentum complexes are quasilocal
expressions for energy-momentum. They reached the conclusion that these
pseudotensorial definitions and the quasilocal expressions are connected in a
direct manner, and that every energy-momentum complex is associated with a
legitimate Hamiltonian boundary term. Furthermore, each expression for energy
has a geometrical and physical significance due to the connection with the
boundary conditions. All these assumptions emphasize the significance of the
energy-momentum complexes and point out their usefulness for the
energy-momentum localization.

In this paper, using the Moller prescription we calculate the energy
distribution of the charged black holes in generalized dilaton-axion gravity
inspired by low energy string theory.

The remainder of our paper is organized as follows: in Section 2 we present an
overview of the space-time under consideration which describes new black hole
solutions for the Einstein-Maxwell scalar field system inspired by low energy
string theory [21]. These solutions have an electric and a magnetic charge and
some non-rotating black hole solutions are obtained in special limit cases.
The M\o ller energy-momentum complex is described in Section 3. This section
is also devoted to the evaluation of the momenta and energy distributions, and
to the analysis of various aspects of energy. Finally, our concluding remarks
are drawn in Discussion. For our calculations we consider the signature
($1,-1,-1,-1$), geometrized units ($c=1$;$G=1$) and assume that Greek (Latin)
indices take value from $0$ to $3$ ($1$ to $3$).

\section{Charged Black Holes Generated in Einstein-Maxwell-Dilaton-Axion
Theory}

Recently, S. Sur, S. Das and S. SenGupta [21] have discovered new black hole
solutions for Einstein-Maxwell scalar field system inspired by low energy
string theory. They considered the action in which two scalar fields are
minimally coupled to Einstein-Hilbert-Maxwell field in the Einstein frame in
four dimension as
\begin{equation}
I=\frac{1}{2\kappa}\int d^{4}x\sqrt{-g}\left[  R-\frac{1}{2}\partial_{\mu
}\varphi\partial^{\mu}\phi-\frac{\omega(\phi)}{2}\partial_{\mu}\xi
\partial^{\mu}\xi-\alpha(\phi,\xi)F^{2}-\beta(\phi,\xi)F_{\mu\nu}\,^{\ast
}F^{\mu\nu}\right]  ,\tag{1}%
\end{equation}
where $\kappa=8\pi G$, $R$ represents the curvature scalar, $F_{\mu\nu}$ is
the Maxwell field tensor, $F$ is the contracted Maxwell scalar i.e. $F_{\mu
}^{\mu}=F$ while $\phi$ and $\xi$ are two massless scalar or pseudo scalar
fields which are coupled to Maxwell field with the functional relationship
$\alpha$ and $\beta$. Here, $\xi$ acquires a non minimal kinetic term $\omega
$. In the context of low energy string theory, the fields $\phi$ and $\xi$ can
be identified as massless scalar dilaton and pseudoscalar axion, respectively.
Two other important quantities are the effective scalar field $\psi(r)$ that
is defined in terms of $\phi$ and $\xi$ as $\psi^{^{\prime}2}=\phi^{^{\prime
}2}+\omega\,\xi^{^{\prime}2}$, and the effective coupling $\gamma(r)$. Sur et
al [21] have found a most general class of static spherically symmetric black
hole solutions classified as asymptotically flat and asymptotically non-flat
types (Section 4 in [21]).

Considering a generalized form of the above action in (1), with the
corresponding connections $\omega(\phi)=e^{2a\phi}$, $\alpha(\phi)=e^{-a\phi}$
and $\beta(\xi)=b\xi$ where $a$ is a real constant which is also non-negative,
Sur et al [21] have analyzed their solutions in the context of the low energy
effective string theory (Section 5 in [21]).

We present the asymptotically flat and the asymptotically non-flat\ black
holes solutions obtained by Sur et al [21] and which are in general
ellectrically and magnetically charged.

For asymptotically flat black holes the metric is given by
\begin{equation}
ds^{2}=f(r)dt^{2}-f(r)^{-1}dr^{2}-h(r)(d\theta^{2}+\sin^{2}\theta d\varphi
^{2}),\tag{2}%
\end{equation}
where
\begin{equation}
f(r)=\frac{(r-r_{-})(r-r_{+})}{(r-r_{0})^{(2-2n)}(r+r_{0})^{2n}}\tag{3}%
\end{equation}
and%
\begin{equation}
h(r)=\frac{(r+r_{0})^{2n}}{(r-r_{0})^{(2n-2)}},\tag{4}%
\end{equation}
whith $0<n<1$ and $r_{0}$ is a constant real parameter.

Also other various parameters are given by%

\begin{align}
r_{\pm}  &  =m_{0}\pm\sqrt{m_{0}^{2}+r_{0}^{2}-\frac{1}{8}(\frac{K_{1}}%
{n}+\frac{K_{2}}{1-n}}).\tag{5}\\
r_{0}  &  =\frac{1}{16m_{0}}\left(  \frac{K_{1}}{n}-\frac{K_{2}}{1-n}\right)
,\nonumber\\
m_{0}  &  =m-(2n-1)r_{0},\nonumber\\
K_{1}  &  =4n[4r_{0}^{2}+2kr_{0}(r_{+}+r_{-})+k^{2}r_{+}r_{-}],\nonumber\\
K_{2}  &  =4(1-n)r_{+}r_{-},\ \ 0<n<1\nonumber\\
m  &  =\frac{1}{16r_{0}}\left(  \frac{K_{1}}{n}-\frac{K_{2}}{1-n}\right)
+(2n-1)r_{0}.\nonumber
\end{align}

Here $m$ is the mass of the black hole. The effective scalar is defined as%

\begin{equation}
\psi(r)=\psi_{0}+2\sqrt{n(n-1)}\ln(\frac{r-r_{0}}{r+r_{0}}) \tag{6}%
\end{equation}

and the effective coupling is given by%

\begin{equation}
\gamma(r)=K_{1}(\frac{r-r_{0}}{r+r_{0}})^{2(1-n)}+K_{2}\,(\frac{r-r_{0}%
}{r+r_{0}})^{-2n}. \tag{7}%
\end{equation}

After performing some calculations the total (bare) electric and magnetic
charges $Q_{e}$ and $Q_{m}$ are found to be connected to the scalar field
shielded electric and magnetic charges $q_{e}$ and $q_{m}$ through the relations%

\begin{equation}
Q_{e}=(q_{e}-q_{m}\,b\,\xi_{0})e^{\alpha\phi_{0}},\;Q_{m}=q_{m} \tag{8}%
\end{equation}
and the the electromagnetic field strengths $F_{tr}$ and $F_{\theta\varphi}$
are given by%

\begin{equation}
F_{tr}=\frac{[Q_{e}e^{-\alpha\phi_{0}}-Q_{m}\,b(\xi-\xi_{0})]e^{\alpha\phi}%
}{(r-r_{0})^{2(1-n)}(r+r_{0})^{2n}}dt\wedge dr,\;F_{\theta\varphi}=Q_{m}%
\,\sin\theta\,d\theta\wedge d\varphi. \tag{9}%
\end{equation}

The asymptotically non-flat black holes are obtained for%

\begin{equation}
f(r)=\frac{(r-r_{-})(r-r_{+})}{r^{2}(2r_{0}/r)^{2n}}, \tag{10}%
\end{equation}

\begin{equation}
h(r)=r^{2}(\frac{2r_{0}}{r})^{2n}, \tag{11}%
\end{equation}

\begin{equation}
r_{\pm}=(\frac{1}{1-n})[m\pm\sqrt{m^{2}-(1-n)\frac{K^{2}}{4}}] \tag{12}%
\end{equation}

and for the $\psi(r)$ and $\gamma(r)$ given by%

\begin{equation}
\psi(r)=\psi_{0}-2\sqrt{n(n-1)}\ln(\frac{2\,r_{0}}{r}), \tag{13}%
\end{equation}

\begin{equation}
\gamma(r)=(4\,n\,r^{2}+K_{2})(\frac{2\,r_{0}}{r})^{2n}. \tag{14}%
\end{equation}

The presence of the parameters $a$ and $b$ in the generalized action for
Einstein-Maxwell theory in four dimensions, coupled to the massless scalar
dilaton $\phi$ and the massless pseudoscalar axion $\xi$ in Einstein frame has
two motivations. The role of the parameter $a$ is to be a regulator for the
strength of the coupling between the dilaton and the Maxwell field. The
parameter $b$ is connected with the Kalb-Ramond tensor $H_{\mu\nu\lambda}$
which appears in the four dimensional heterotic string action [21] (see eq.
5.3 therein). Another explanation for the introduction of the parameters $a$
and $b$ is that for some specific values the generalized action (see eq. 5.1
in [21]) yields the field equations which correspond to a four dimensional
effective compactified version of a higher dimensional (bulk)
Einstein-Maxwell-Kalb-Ramond theory in a Randall-Sundrum scenario that is
connected to the Planck-electroweak hierarchy problem. Some particular values
of the parameters $a$ and $b$ lead to special cases, for $a=1$ the field
theoretic limit in the case of the ten dimensional or the effective four
dimensional superstring model, in the bosonic sector is reached. For
$a=\sqrt{1+2/n}$ the four dimensional Kaluza-Klein toroidal reduction of a
$4+n$ dimensional theory is recovered. The case of usual Einstein-Maxwell
theory that is coupled minimally with a massless Klein-Gordon scalar field
$\phi$ is obtained for $a=0$ ignoring the presence of the other scalar $\xi$
(or the KR tensor $H_{\mu\nu\lambda}$).

The effective field equations obtained for the general formalism take a new
form. Solving these equations for the asymptotically flat and asymptotically
non-flat black holes and imposing some specific values for the parameters $a$
and $b$ the expressions for $\phi(r)$ and $\xi(r)$ are determined.

In this paper we evaluate the energy and momentum distributions in the
M\o ller prescription for asymptotically flat (AF) and asymptotically non-flat
(ANF) solutions in the context of low energy string theory. Taking into
account two special values as $\left\vert b\right\vert =\left\vert
a\right\vert $ and $\left\vert b\right\vert \neq\left\vert a\right\vert $ with
some particular cases for the parameters $a$ and $b$ we also analyze various
aspects of the energy distribution.

\section{Energy and Momentum in the M\o ller Prescription}

We perform the calculations in the M\o ller prescription in the Einstein frame
applying this definition to the metrics given by (2), (3), (4), (10) and (11)
because we don't need to carry out the calculations in quasi-Cartesian
coordinates. Next, we briefly revise the expressions for
the\ M\o ller\ energy-momentum complex $\mathcal{\Theta}_{\nu}^{\mu}$, the
M\o ller superpotential $M_{\nu}^{\mu\lambda}$, the energy density
$\mathcal{\Theta}_{0}^{0}$ and the momentum density $\mathcal{\Theta}_{i}^{0}$
components, and also the expressions\ for the energy and momentum $P_{\mu}$.

The M\o ller energy-momentum complex [11] is given by the definition%

\begin{equation}
\mathcal{\Theta}_{\nu}^{\mu}=\frac{1}{8\pi}M_{\nu\,\,,\,\lambda}^{\mu\lambda
},\tag{15}%
\end{equation}
where $M_{\nu}^{\mu\lambda}$ represents M{\o }ller's superpotential%
\begin{equation}
M_{\nu}^{\mu\lambda}=\sqrt{-g}\left(  \frac{\partial g_{\nu\sigma}}{\partial
x^{\kappa}}-\frac{\partial g_{\nu\kappa}}{\partial x^{\sigma}}\right)
g^{\mu\kappa}g^{\lambda\sigma}.\tag{16}%
\end{equation}
\qquad\qquad The M{\o }ller superpotential is antisymmetric
\begin{equation}
M_{\nu}^{\mu\lambda}=-M_{\nu}^{\lambda\mu}.\tag{17}%
\end{equation}
\qquad The M{\o }ller energy-momentum complex holds the local conservation
law
\begin{equation}
\frac{\partial\mathcal{\Theta}_{\nu}^{\mu}}{\partial x^{\mu}}=0,\tag{18}%
\end{equation}
where $\mathcal{\Theta}_{0}^{0}$ and $\mathcal{\Theta}_{i}^{0}$ represent the
energy density and and the momentum density components, respectively.

The energy and momentum are given by
\begin{equation}
P_{\mu}=\int\int\int_{\mu}^{0}\mathcal{\Theta}_{\mu}^{0}dx^{1}dx^{2}%
dx^{3}.\tag{19}%
\end{equation}

For the metric given by (2) the components of the M{\o }ller superpotential
have the following expressions%

\begin{equation}
M_{0}^{01}=h(r)\frac{\partial f(r)}{\partial r}\sin\theta, \tag{20}%
\end{equation}

\begin{equation}
M_{2}^{21}=f(r)\frac{\partial h(r)}{\partial r}\sin\theta, \tag{21}%
\end{equation}

\begin{equation}
M_{3}^{31}=f(r)\frac{\partial h(r)}{\partial r}\sin\theta, \tag{22}%
\end{equation}

\begin{equation}
M_{3}^{32}=2\,\cos\theta. \tag{23}%
\end{equation}

The equations (20)-(23) present a dependence on the metric functions $f(r)$
and $h(r)$, on their first derivative with respect to $r$ coordinate
$\frac{\partial f(r)}{\partial r}$ and $\frac{\partial h(r)}{\partial r}$, and
on $\theta$ coordinate through $\sin\theta$ and $\cos\theta$. The expression
for energy in the case of a nonstatic spherically symmetric metrics was
calculated in [17] (see, in particular Astrophys. Space. Sci. 283, 23 (2003)).
For the metrics described by (2)-(4) and (2), (10), (11) all the momenta
vanish. Using (19) and (20) we can calculate the expressions for energy.

We return to the asymptotically flat and asymptotically non-flat black hole
solutions and perform our study considering the special values $\left\vert
b\right\vert =\left\vert a\right\vert $ and $\left\vert b\right\vert
\neq\left\vert a\right\vert $ and some particular cases for the parameters $a$
and $b$. In the asymptotic limit the connections between $\phi$, $\xi$,
$\phi^{^{\prime}}$, $\xi^{^{\prime}}$, $K_{1}$, $K_{2}$, $q_{e}$, $q_{m}$,
$Q_{e}$, $Q_{m}$, $a$, $b$, $r$, $r_{+}$, $r_{-}$, $r_{0}$ and $Q^{2}=$
$Q_{e}^{2}+Q_{m}^{2}$ are given in [21] (see equations 5.14 and 5.15 therein).

1) Firstly, we present the results for the asymptotically flat black hole solutions.

Case I. $\left\vert b\right\vert =\left\vert a\right\vert $

The eqs. 5.15 in [21] are satisfied uniquely for the values $n=1/(1+a^{2})$
and $K_{2}=0$, leading to the following expressions for $r_{0}$, $m_{0}$,
$r_{+}$ and $r_{-}$%

\begin{equation}
r_{0}=\frac{(1+a^{2})Q^{2}\,e^{-\alpha\phi_{0}}}{4\,m_{0}}, \tag{24}%
\end{equation}

\begin{equation}
m_{0}=m-\frac{(1-a^{2})}{(1+a^{2})}r_{0},\;(Q^{2}=Q_{e}^{2}+Q_{m}^{2})
\tag{25}%
\end{equation}

\begin{equation}
r_{+}=2\,m_{0}-r_{0}, \tag{26}%
\end{equation}

\begin{equation}
r_{-}=r_{0}. \tag{27}%
\end{equation}

Performing a coordinate shift $r+r_{0}\rightarrow r$ the metric described by
(2), (3) and (4) can be written in a new form%

\begin{align}
ds^{2}  &  =(1-\frac{2m_{0}}{r})(1-\frac{2r_{0}}{r})^{\frac{1-a^{2}}{1+a^{2}}%
}\,dt^{2}-(1-\frac{2m_{0}}{r})^{-1}(1-\frac{2r_{0}}{r})^{\frac{a^{2}-1}%
{a^{2}+1}}\,dr^{2}-\tag{28}\\
&  -r^{2}(1-\frac{2r_{0}}{r})^{\frac{2\,a^{2}}{1+a^{2}}}\,(d\theta^{2}%
+\sin^{2}\theta d\varphi^{2}).\nonumber
\end{align}

The dilaton field $\phi(r)$, the axion field $\xi(r)$ and the electromagnetic
field strengths $F_{tr}$ and $F_{\theta\varphi}$ are expressed by equations
5.20 and 5.21 of [21] with $r_{0}$ and $m_{0}$ given by (24) and (25).

Using (15) and (19) we obtain that the expression for energy in the M\o ller
prescription is given by%

\begin{equation}
E(r)=\frac{m_{0}\,a^{2}\,r+m_{0}\,r-4\,m_{0}\,r_{0}-r_{0}\,a^{2}\,r+r_{0}%
\,r}{r(a^{2}+1)}. \tag{29}%
\end{equation}

From (29) we notice that the energy distribution depends on the parameters
$m_{0}$, $a$, $r_{0}$ and $r$.

There are 3 particular limiting cases that we present in the following.

a. For $a=b=1$ we lead to the bosonic sector of the ten dimensional heterotic
superstring toroidally compactified to four spacetime dimensions. The metric
given by (28) and the dilaton and axion fields have a new form [21] (see
equations 5.23 and 5.24 therein) and the energy is%

\begin{equation}
E(r)=m(1-\frac{2r_{0}}{r})=m-\frac{Q^{2}\,e^{-\phi_{0}}}{r}. \tag{30}%
\end{equation}

where $r_{0}=Q^{2}\,e^{-\phi_{0}}/(2\,m)$. If $Q_{e}=0$, $Q_{m}=Q$ or
$Q_{m}=0$, $Q_{e}=Q$ we recover the solutions given by Garfinkle, Horowitz and
Strominger (GHS) [22] and Gibbons [23] and explained by Gibbons and Maeda (GM)
[24] (the solutions are elaborated in [22] and [23] assuming a zero value or
at least a trivial value for the KR axion field). These solutions describe a
magnetically or electrically charged dilaton black hole. The non-trivial
dilaton-axion configuration can be obtained using a magnetically (or,
electrically) charged dilaton black hole configuration with the help of the
SL(2,R) invariance, even when the value of the parameter $a\neq1$.

b. In the case $a=b<<1$, after some calculations [21] (solving equation 5.20
therein using $r_{0}=Q^{2}/(4\,m_{0})+O(a^{2})$), is demonstrated that the
black hole solutions are characterized by the parameters%

\begin{equation}
\phi(r)=\phi_{0}+\frac{4\,a\,r_{0}}{r}(\frac{Q_{m}^{2}-Q_{e}^{2})}{Q^{2}%
}+O(a^{3}),\;\xi(r)=\xi_{0}+\frac{4\,a\,r_{0}}{r}(\frac{Q_{m}\,Q_{e}}{Q^{2}%
})+O(a^{3}), \tag{31}%
\end{equation}

\begin{equation}
r_{0}=\frac{1}{2}(m-\sqrt{m^{2}-Q^{2}})+O(a^{2}),\;m_{0}=\frac{1}{2}%
(m+\sqrt{m^{2}-Q^{2}})+O(a^{2}). \tag{32}%
\end{equation}

For $a\rightarrow0$ this is the case of the standard dyonic
Reissner-Nordst\"{o}m black hole solution. Using (29) and (32) the energy becomes%

\begin{equation}
E(r)=m-\frac{Q^{2}}{r}. \tag{33}%
\end{equation}

c. For $a=b>>1$ the parameters $r_{0}$ and $m_{0}$ have the expressions [21]%

\begin{equation}
r_{0}\approx\frac{a^{2}\,Q^{2}\,e^{-\alpha\phi_{0}}}{4\,m_{0}},\;m_{0}\approx
m+r_{0}. \tag{34}%
\end{equation}

Considering that in the limit $a\rightarrow\infty$ the constants $r_{0}$ and
$m_{0}$ could not be larger than $m$ and after some calculations the dilaton
and axion fields, respectively are given by eqs. 5.29 in [21] and the metric is%

\begin{align}
ds^{2}  &  =(1-\frac{2\,m}{r-2\,r_{0}})dt^{2}-(1-\frac{2\,m}{r-2\,r_{0}}%
)^{-1}dr^{2}-\tag{35}\\
&  -(r-2\,r_{0})^{2}(d\theta^{2}+\sin^{2}\theta d\varphi^{2}).\nonumber
\end{align}

With a coordinate changing in $r-2\,r_{0}=r$ the standard Schwarzschild black
hole is obtained together with non-zero solutions for the dilaton, axion and
the $U(1)$ gauge field.

The expression for energy is given by%

\begin{equation}
E=m. \tag{36}%
\end{equation}

This expression also represents\ the ADM mass of the black hole.

Case II. $\left\vert b\right\vert \neq\left\vert a\right\vert $

As is demonstrated in [21], in this situation it is not always possible to
construct an analytic closed form black hole solution from the given metric
ansatz, as only some special values enable this scheme. For the string theory
the case $a=1$ and $b<<1$ is of importance, and the axion field $\xi$ is
trivial up to $O(b)$ (equation 5.32 in [21]). The metric has a new form%

\begin{align}
ds^{2}  &  =\frac{(r-r_{+})(r-r_{-})}{r^{2}-r_{0}^{2}}dt^{2}-\frac{r^{2}%
-r_{0}^{2}}{(r-r_{+})(r-r_{-})}dr^{2}-\tag{37}\\
&  -(r^{2}-r_{0}^{2})(d\theta^{2}+\sin^{2}\theta d\varphi^{2})\nonumber
\end{align}

and is described by the dilaton charge $Q_{\phi}=\frac{(Q_{m}^{2}-Q_{e}%
^{2})\,e^{-\phi_{0}}}{m}$\ and by the quantities%

\begin{equation}
\phi(r)=\phi_{0}+\ln(\frac{r-r_{0}}{r+r_{0}}),\;\xi(r)=\xi_{0}, \tag{38}%
\end{equation}

\begin{equation}
F_{tr}=\frac{Q_{e}}{(r+r_{0})^{2}}dt\wedge dr,\;F_{\theta\varphi}=Q_{m}%
\,\sin\theta\,d\theta\wedge d\varphi, \tag{39}%
\end{equation}

with%

\begin{equation}
r_{0}=\frac{(Q_{e}^{2}-Q_{m}^{2})\,e^{-\phi_{0}}}{2\,m},\;r_{\pm}=m\pm
\sqrt{m^{2}+r_{0}^{2}-(Q_{e}^{2}+Q_{m}^{2})e^{-\phi_{0}}}. \tag{40}%
\end{equation}

This black hole solution presents two horizons and two charges, electric and magnetic.

The calculations performed with (15), (19) applied to (37) yield the energy in
the M\o ller prescription%

\begin{equation}
E(r)=\frac{r_{-}r^{2}+r_{+}r^{2}+r_{-}r_{0}^{2}+r_{+}r_{0}^{2}-2\,r\,r_{0}%
^{2}-2\,r\,r_{+}r_{-}}{2(r^{2}-r_{0}^{2})}. \tag{41}%
\end{equation}

From (38), (39), (40) and (41) it results that the energy distribution depends
on the mass $m$, the total (bare) electric and magnetic charges $Q_{e}$ and
$Q_{m}$, $r$ and $\phi_{0}$.

Using (40) in (41) we obtain%

\begin{equation}
E(r)=\frac{8\,m^{3}\,r^{2}+2\,m(Q_{e}^{2}-Q_{m}^{2})^{2}e^{-2\,\phi_{0}%
}-8\,m^{2}\,r(Q_{e}^{2}+Q_{m}^{2})e^{-\phi_{0}}}{2[4\,m^{2}\,r^{2}-(Q_{e}%
^{2}-Q_{m}^{2})^{2}e^{-2\,\phi_{0}}]}. \tag{42}%
\end{equation}

In the special case of $Q_{e}=0$ or $Q_{m}=0$ combined with the coordinate
transformation $r+r_{0}\rightarrow r$\ the (GHS) [22] magnetically or
electrically charged black holes are recovered [21]. For $Q_{e}=Q_{m}$ or
$Q_{e}=-Q_{m}$ one gets a vanishing value for the dilaton charge and this
leads to the case of the standard Reissner-Nordst\"{o}m black hole solution [21].

2) We consider now the asymptotically non-flat black hole solutions and we
calculate the energy in the M\o ller prescription for the same cases
$\left\vert b\right\vert =\left\vert a\right\vert $ and $\left\vert
b\right\vert \neq\left\vert a\right\vert $ considering some special values.

Case I. $\left\vert b\right\vert =\left\vert a\right\vert $

The solutions are given by the value $n=1/(1+a^{2})$ and $K_{2}=0$, with
$r_{+}=2\,m/(1+n)$ and $r_{-}=0$. The metric is given by%

\begin{align}
ds^{2}  &  =(\frac{r}{2r_{0}})^{2n}[1-\frac{2\,m}{(1-n)r}]dt^{2}-(\frac
{2r_{0}}{r})^{2n}[1-\frac{2\,m}{(1-n)r}]^{-1}dr^{2}-\tag{43}\\
&  -r^{2}(\frac{2r_{0}}{r})^{2n}(d\theta^{2}+\sin^{2}\theta d\varphi
^{2}).\nonumber
\end{align}

The dilaton field, the axion field and the electromagnetic field strengths
$F_{tr}$ and $F_{\theta\varphi}$ are given by eqs. 5.45 and 5.46 in [21], with
$q^{2}=q_{e}^{2}+q_{m}^{2}$. This describes a black hole with a causal
structure similar to the standard Schwarzschild black hole and with a null
hypersurface obtained for $r=2\,m/(1-n)$, which is also the value for which
the event horizon is regular.

a. In the case $a=b=1$ in the Einstein frame the metric has the form%

\begin{equation}
ds^{2}=(\frac{r-4\,m}{2\,r_{0}})dt^{2}-(\frac{2\,r_{0}}{r-4\,m})dr^{2}%
-2\,r_{0}\,r(d\theta^{2}+\sin^{2}\theta d\varphi^{2}), \tag{44}%
\end{equation}

with the dilaton and axion fields given in equations 5.48 in [21] and with the
electromagnetic field strengths $F_{tr}=$ $q_{e}/(2\,q^{2})dt\wedge dr$ and
$F_{\theta\varphi}=q_{m}\,\sin\theta\,d\theta\wedge d\varphi$. In the special
cases $q_{e}=0$ or $q_{m}=0$ the axion field $\xi$ vanishes and we recover the
cases of magnetically or electrically charged dilaton black holes with curved
asymptotes [25]. We also notice that asymptotically non-flat magnetically
charged dilaton black hole solutions for particular values of the mass $m$ and
magnetic charge $q_{m}$ have been developed [26].

The expression for energy calculated with the M\o ller definition is given by%

\begin{equation}
E(r)=\frac{r}{2}. \tag{45}%
\end{equation}

b. For $a=b<<1$ we have $n=1/(1+a^{2})\approx1$ and the metric becomes%

\begin{align}
ds^{2}  &  =(\frac{r}{2r_{0}})^{2}[1-\frac{2\,m}{a^{2}\,r}]dt^{2}%
-(\frac{2r_{0}}{r})^{2}[1-\frac{2\,m}{a^{2}\,r}]^{-1}dr^{2}-\tag{46}\\
&  -4\,r_{0}^{2}(d\theta^{2}+\sin^{2}\theta d\varphi^{2}),\nonumber
\end{align}

where the dilaton and axion fields are expressed with equations 5.54 in [21].

For the energy distribution we obtain%

\begin{equation}
E(r)=\frac{1}{2}\{\frac{r}{2r_{0}^{2}}[1-\frac{2\,m}{a^{2}\,r}]+\frac
{m}{2\,a^{2}\,r_{0}^{2}}\}4\,r_{0}^{2}=r-\frac{m}{a^{2}}. \tag{47}%
\end{equation}

The energy distribution presents a dependence on $r_{0}$, $r$, the mass $m$ of
the black hole and the parameter $a$. This expression can be also written%

\begin{equation}
E(r)=\frac{1}{2}\{2r[1-\frac{2\,m}{r}]+2m\}. \tag{48}%
\end{equation}

In the limit $a\rightarrow0$ the expression for energy given by (47) diverges.
At large distances $r\rightarrow\infty$ the energy distribution tends toward infinity.

c. In the case $a=b>>1$ with $n=1/(1+a^{2})\approx1/a^{2}$ the solution is
described by the metric%

\begin{align}
ds^{2}  &  =(\frac{r}{2r_{0}})^{\frac{2}{a^{2}}}[1-\frac{2\,a^{2}\,m}%
{(a^{2}-1)r}]dt^{2}-(\frac{2r_{0}}{r})^{\frac{2}{a^{2}}}[1-\frac{2\,a^{2}%
\,m}{(a^{2}-1)r}]^{-1}dr^{2}-\tag{49}\\
&  -r^{2}(\frac{2r_{0}}{r})^{\frac{2}{a^{2}}}(d\theta^{2}+\sin^{2}\theta
d\varphi^{2}),\nonumber
\end{align}

with the dilaton and axion fields given in equations 5.56 in [21] and with the
electromagnetic field strengths $F_{tr}\approx$ $q_{e}/(a^{2}\,q^{2})dt\wedge
dr$ and $F_{\theta\varphi}=q_{m}\,\sin\theta\,d\theta\wedge d\varphi$.

The corresponding calculations using (15), (19) and (49) lead to the
expression for energy which is given by%

\begin{equation}
E(r)=\frac{r\,a^{2}-r-2\,a^{2}\,m+a^{4}\,m}{(a^{2}-1)a^{2}}. \tag{50}%
\end{equation}

In the limit $a\rightarrow\infty$ we recover the energy for the Schwarzschild
black hole solution%

\begin{equation}
E=m. \tag{51}%
\end{equation}

As in the case $a=b>>1$ for the asymptotically flat black hole solutions this
expression also represents the ADM mass of the black hole, even if the
solution is non-flat asymptotically for a finite value of the parameter $a$.

Case II. $\left\vert b\right\vert \neq\left\vert a\right\vert $

Like in the case of the asymptotically flat black hole solutions is not
allowed to develop analytic closed form black hole solutions. We have to take
into account special values for the parameters $a$ and $b$, $a=1$ and $b<<1$
and consider the axion $\xi$ trivial up to order $b$ and written as $\xi
=\xi_{0}+O(b)$. The solution is developed neglecting the $O(b)$ terms and for
$n=1/2$ and $K_{2}=2\,q_{e}^{2}\,q_{m}^{2}/r_{0}^{2}$. The metric corresponds
to a dyonic black hole given by%

\begin{align}
ds^{2}  &  =\frac{(r-r_{+})(r-r_{-})}{2\,r_{0}\,r}dt^{2}-\frac{2\,r_{0}%
\,r}{(r-r_{+})(r-r_{-})}dr^{2}-\tag{52}\\
&  -2\,r_{0}\,r(d\theta^{2}+\sin^{2}\theta d\varphi^{2}),\nonumber
\end{align}

described by the quantities%

\begin{equation}
\phi(r)=-\ln(\frac{q_{e}^{2}}{r_{0}\,r}),\;\xi(r)=\xi_{0}, \tag{53}%
\end{equation}

\begin{equation}
F_{tr}=\frac{1}{2\,q_{e}}dt\wedge dr,\;F_{\theta\varphi}=q_{m}\,\sin
\theta\,d\theta\wedge d\varphi, \tag{54}%
\end{equation}

where%

\begin{equation}
r_{\pm}=2(m\pm\sqrt{m^{2}-\frac{q_{e}^{2}\,q_{m}^{2}}{4\,r_{0}^{2}}}).
\tag{55}%
\end{equation}

The energy in the M\o ller prescription is given by%

\begin{equation}
E(r)=\frac{1}{2}\frac{r^{2}-r_{+}r_{-}}{r}=\frac{1}{2}\frac{r^{2}-\frac
{q_{e}^{2}\,q_{m}^{2}\,}{r_{0}^{2}}}{r}. \tag{56}%
\end{equation}

In this case the expression for energy depends explicitly on the
electromagnetic charges $q_{e}$ and $q_{m}$. It is interesting to notice that
in this case the interchange of $q_{e}$ and $q_{m}$ does not modify the
expression for energy. In the limit cases $r\rightarrow0$ and $r\rightarrow
\infty$ the enegy distribution diverges.

\section{Discussion}

We calculate the energy and momentum distributions in the M\o ller
prescription for some asymptotically flat and asymptotically non-flat
solutions in the context of Einstein-Maxwell-dilaton-axion theory [21]. It is
important to emphasize that using the M\o ller energy-momentum complex the
requirement of performing the calculations in quasi-Cartesian coordinates can
be avoided.

We consider two special cases $\left\vert b\right\vert =\left\vert
a\right\vert $ and $\left\vert b\right\vert \neq\left\vert a\right\vert $. For
$\left\vert b\right\vert =\left\vert a\right\vert $ we\ investigate the
corresponding three special values for the parameters $a$ and $b$, which are
$a=b=1$, $a=b<<1$ and $a=b>>1$ and some limit cases. In the case $\left\vert
b\right\vert \neq\left\vert a\right\vert $ the special values $a=b<<1$ yield
important results and also some particular cases are studied. For all the
particular cases mentioned the momenta are found to be zero.

In the Table 1 and Table 2 we briefly present the expressions for energy
obtained in the case of asymptotically flat and asymptotically non-flat
solutions, respectively, and some limit cases that occur in each situation.

Firstly, we outline the results for the case of asymptotically flat black hole solutions.%

\[%
\begin{tabular}
[c]{ll}%
Case & Energy distribution\\
$\left\vert b\right\vert =\left\vert a\right\vert $ & $E(r)=\frac{m_{0}%
\,a^{2}\,r+m_{0}\,r-4\,m_{0}\,r_{0}-r_{0}\,a^{2}\,r+r_{0}\,r}{r(a^{2}+1)}$\\
$\left\vert b\right\vert =\left\vert a\right\vert $,\thinspace\ $a=b=1$ &
$E(r)=m(1-\frac{2r_{0}}{r})=m-\frac{Q^{2}\,e^{-\phi_{0}}}{r}$\\
$\left\vert b\right\vert =\left\vert a\right\vert $,\thinspace\ $a=b=1$,
$Q_{e}=0$, $Q_{m}=Q$ & $E(r)=m-\frac{Q_{m}^{2}\,e^{-\phi_{0}}}{r}$ (GHS)\\
$\left\vert b\right\vert =\left\vert a\right\vert $,\thinspace\ $a=b=1$,
$Q_{m}=0$, $Q_{e}=Q$ & $E(r)=m-\frac{Q_{e}^{2}\,e^{-\phi_{0}}}{r}$ (GHS)\\
$\left\vert b\right\vert =\left\vert a\right\vert $, $a=b<<1$, limit case
$a\rightarrow0$ & $E(r)=m-\frac{Q^{2}}{r}$ standard dyonic RN black hole\\
$\left\vert b\right\vert =\left\vert a\right\vert $, $a=b>>1$, limit
case$\ a\rightarrow\infty$ & $E=m$ standard Schwarzschild black hole\\
$\left\vert b\right\vert \neq\left\vert a\right\vert $, $a=1$ and $b<<1$ &
$E(r)=\frac{8\,m^{3}\,r^{2}+2\,m(Q_{e}^{2}-Q_{m}^{2})^{2}e^{-2\,\phi_{0}%
}-8\,m^{2}\,r(Q_{e}^{2}+Q_{m}^{2})e^{-\phi_{0}}}{2[4\,m^{2}\,r^{2}-(Q_{e}%
^{2}-Q_{m}^{2})^{2}e^{-2\,\phi_{0}}]}$\\
$\left\vert b\right\vert \neq\left\vert a\right\vert $, $a=1$ and $b<<1$,
$Q_{e}=0$ & $E(r)=\frac{8\,m^{3}\,r^{2}+2\,mQ_{m}^{4}e^{-2\,\phi_{0}}%
-8\,m^{2}\,rQ_{m}^{2}e^{-\phi_{0}}}{2[4\,m^{2}\,r^{2}-Q_{m}^{4}e^{-2\,\phi
_{0}}]}$ (GHS)\\
$\left\vert b\right\vert \neq\left\vert a\right\vert $, $a=1$ and $b<<1$,
$Q_{m}=0$ & $E(r)=\frac{8\,m^{3}\,r^{2}+2\,mQ_{e}^{4}e^{-2\,\phi_{0}}%
-8\,m^{2}\,rQ_{e}^{2}e^{-\phi_{0}}}{2[4\,m^{2}\,r^{2}-Q_{e}^{4}e^{-2\,\phi
_{0}}]}$ (GHS)\\
$\left\vert b\right\vert \neq\left\vert a\right\vert $, $a=1$ and $b<<1$,
$Q_{e}=Q_{m}$ or $Q_{e}=-Q_{m}$ & $E(r)=m-\frac{2Q^{2}e^{-\phi_{0}}}{r}$
standard RN black hole
\end{tabular}
\ \ \ \
\]

\[
\text{Table 1}%
\]

Now, we point out the results obtained for the asymptotically non-flat black
hole solutions.%

\[%
\begin{tabular}
[c]{ll}%
Case & Energy distribution\\
$\left\vert b\right\vert =\left\vert a\right\vert $,\thinspace\ $a=b=1$ &
$E(r)=\frac{r}{2}$\\
$\left\vert b\right\vert =\left\vert a\right\vert $, $a=b<<1$ & $E(r)=r-\frac
{m}{a^{2}}$\\
$\left\vert b\right\vert =\left\vert a\right\vert $, $a=b<<1$, limit case
$a\rightarrow0$ & $E(r)\rightarrow-\infty$\\
$\left\vert b\right\vert =\left\vert a\right\vert $, $a=b<<1$, limit case
$r\rightarrow\infty$ & $E(r)\rightarrow\infty$\\
$\left\vert b\right\vert =\left\vert a\right\vert $, $a=b>>1$ & $E(r)=\frac
{r\,a^{2}-r-2\,a^{2}\,m+a^{4}\,m}{(a^{2}-1)a^{2}}$\\
$\left\vert b\right\vert =\left\vert a\right\vert $, $a=b>>1$, limit case
$a\rightarrow\infty$ & $E=m$ standard Schwarzschild black hole\\
$\left\vert b\right\vert \neq\left\vert a\right\vert $, $a=1$ and $b<<1$ &
$E(r)=\frac{1}{2}\frac{r^{2}-r_{+}r_{-}}{r}=\frac{1}{2}\frac{r^{2}-\frac
{q_{e}^{2}\,q_{m}^{2}\,}{r_{0}^{2}}}{r}$\\
$\left\vert b\right\vert \neq\left\vert a\right\vert $, $a=1$ and $b<<1$,
limit case $r\rightarrow0$ & $E(r)\rightarrow\pm\infty$\\
$\left\vert b\right\vert \neq\left\vert a\right\vert $, $a=1$ and $b<<1$,
limit case $r\rightarrow\infty$ & $E(r)\rightarrow\infty$%
\end{tabular}
\ \ \ \
\]

\[
\text{Table 2}%
\]

The expression for energy $E(r)=m-\frac{Q^{2}}{r}$ obtained in the case of
asymptotically flat black hole solutions for $\left\vert b\right\vert
=\left\vert a\right\vert $, $a=b<<1$, limit case $a\rightarrow0$ and
$\left\vert b\right\vert \neq\left\vert a\right\vert $, $a=1$ and $b<<1$,
$Q_{e}=Q_{m}$ or $Q_{e}=-Q_{m}$, respectively is in good agreement with the
result given by Komar [27].

All these results illustrate that the use of the M\o ller prescription for the
evaluation of the expressions for energy is an important option. We notice
that interesting particular cases arise for both classes of solutions AF and
ANF, respectively.

For future work, we intend to explore the results yielded by the
pseudotensorial method for these black hole solutions using other
energy-momentum complexes.


\begin{thebibliography}{99}                                                                                               %


\bibitem {1}L. Bel, C. R Acad. Sci. Paris, \textbf{246}, 3105, (1958); L. Bel,
C. R. Acad. Sci., Paris \textbf{248}, 1292 (1959); I. Robinson, unpublished
lectures, Kings College, London (1958); I. Robinson, Class. Quant Grav.
\textbf{14}, 4331, (1997); M. Chevreton, Nuovo Cimento \textbf{34}, 90 (1964);
J. M. M. Senovilla, Class. Quant Grav. \textbf{17}, 2799 (2000); J. M. M.
Senovilla, Gravitation and Relativity in General, eds. A. Molina, J. Mart%
\'{}%
\i n, E. Ruiz and F. Atrio (World Scientific, 1999), gr-qc/9901019; G.
Bergqvist, Commun. Math. Phys. \textbf{207}, 467, (1999); G. Bergqvist, J.
Math. Phys. \textbf{39}, 2141, (1998).

\bibitem {2}G. Bergqvist, I. Eriksson and J. M. M. Senovilla, gr-qc/0303036;
S. Deser, in Gravitation and Relativity in General, eds. A. Molina, J. Mart%
\'{}%
\i n, E. Ruiz and F. Atrio (World Scientific, 1999), gr-qc/9901007; S. Deser,
J. S. Franklin and D. Seminara, Class. Quant Grav. \textbf{16}, 2815, (1999);
S. Deser and D. Seminara, Phys. Rev. \textbf{D62}, 084010 (2000); M. A. G.
Bonilla and J. M. M. Senovilla, Gen. Rel. Grav. \textbf{29}, 91, (1997); J. M.
Pozo and J. M. Parra, Class. Quant. Grav. \textbf{19}, 967 (2002); X. Jaen and
A. Balfagon, Class. Quant. Grav. \textbf{17}, 2491 (2000).

\bibitem {3}J. D. Brown and J. W. York, Quasilocal energy in general
relativity, Mathematical aspects of classical field theory (Seattle, WA,
1991), 129-142, Contemp. Math., 132, Amer. Math. Soc., Providence, RI, 1992;
J. D. Brown and J. W. York, Phys. Rev. \textbf{D47}, 1407 (1993); S. W.
Hawking and G. T. Horowitz, Class. Quantum Grav. \textbf{13}, 1487 (1996);
C-C.M. Liu and S. T. Yau, Phys. Rev. Lett. \textbf{90}, 231102 (2003).

\bibitem {4}T. Shirafuji, G. G. L. Nashed and K. Hayashi, Prog. Theor. Phys.
\textbf{95}, 665 (1996); T. Shirafuji and Gamal G. L. Nashed, Prog. Theor.
Phys. \textbf{98} (1997); Gamal G. L. Nashed, Phys. Rev. \textbf{D66}, 064015
(2002); Gamal G. L. Nashed, Mod. Phys. Lett. \textbf{A22}, 1047 (2007); Gamal
G. L. Nashed and T. Shirafuji, Int. J. Mod. Phys. \textbf{D16}, 65 (2007);
Gamal G. L. Nashed, Eur. Phys. J. \textbf{C49}, 851 (2007); Gamal G. L.
Nashed, Chin. Phys. Lett. \textbf{25}, 1202 (2008); J. W. Maluf, J. Math.
Phys. \textbf{36}, 4242 (1995); J. W. Maluf and A. Kneip, J. Math. Phys.
\textbf{38}, 458 (1997); J. W. Maluf, E. F. Martins and A. Kneip; J. Math.
Phys. \textbf{37}, 6302 (1996); J. W. Maluf, J. Math. Phys. \textbf{37}, 6293
(1996); J. W. Maluf, F. F. Faria and S. C. Ulhoa, Class. Quant. Grav.
\textbf{24}, 2743 (2007); J. W. Maluf, M. V. O. Veiga and J. F. da Rocha-Neto,
Gen. Rel. Grav. \textbf{39}, 227 (2007); Lau Loi So and J. M. Nester,
arXiv:0811.4231; J. M. Nester, Lau Loi So and T. Vargas, Phys. Rev.
\textbf{D78}, 044035 (2008).

\bibitem {5}A. Einstein, Preuss. Akad. Wiss. Berlin \textbf{47}, 778 (1915);
Addendum-ibid. \textbf{47}, 799 (1915); A. Trautman, in Gravitation: an
Introduction to Current Research, ed. L. Witten (Wiley, New York, 1962, p. 169).

\bibitem {6}L. D. Landau and E.M. Lifshitz, The Classical Theory of Fields
(Pergamon Press, 1987, p. 280).

\bibitem {7}A. Papapetrou, Proc. R. Irish. Acad. \textbf{A} \textbf{52}, 11 (1948).

\bibitem {8}P. G. Bergmann and R. Thompson, Phys. Rev. \textbf{89}, 400 (1953).

\bibitem {9}S. Weinberg, Gravitation and Cosmology: Principles and
Applications of General Theory of Relativity (John Wiley and Sons, Inc., New
York, 1972, p. 165).

\bibitem {10}A. Qadir and M. Sharif, Phys. Lett. \textbf{A} \textbf{167}, 331 (1992).

\bibitem {11}C. M\o ller, Ann. Phys. (NY) \textbf{4}, 347 (1958).

\bibitem {12}K. S. Virbhadra, Phys. Rev. \textbf{D41}, 1086 (1990); K. S.
Virbhadra, Phys. Rev. \textbf{D42}, 2919 (1990); N. Rosen and K.S. Virbhadra,
Gen. Rel. Grav. \textbf{25}, 429 (1993); K. S.Virbhadra and J. C. Parikh,
Phys. Lett. \textbf{B331}, 302 (1994); J. M. Aguirregabiria, A. Chamorro and
K. S. Virbhadra, Gen. Rel. Grav. \textbf{28}, 1393 (1996); S. S. Xulu, Int. J.
Theor. Phys. \textbf{37}, 1773 (1998); S. S. Xulu, Int. J. Mod. Phys.
\textbf{D7}, 773 (1998); I. Radinschi, Acta Physica Slovaca, \textbf{49(5)},
789 (1999); I. Radinschi, Mod. Phys. Lett. \textbf{A15}, Nos. 11\&12, 803
(2000); I-Ching Yang and I. Radinschi, Chin. J. Phys. \textbf{41}, 326 (2003);
I. Radinschi and Th. Grammenos, Int. J. Mod. Phys. \textbf{A21}, 2853 (2006);
Th. Grammenos and I. Radinschi, Int. J. Theor. Phys. \textbf{46(4)}, 1055
(2007); T. Bringley, Mod. Phys. Lett. \textbf{A17}, 157 (2002); M. Sukenik and
J. Sima, arXiv:grqc/0101026; M. Sharif, Nuovo Cim. \textbf{B19}, 463 (2004);
P. Halpern, Astrophys. Space. Sci. \textbf{306}, 279 (2006); Amir M. Abbassi,
Saeed Mirshekari, Int. J. Mod. Phys. \textbf{A23}, 4569 (2008); Amir M.
Abbassi, Saeed Mirshekari, Amir H. Abbassi, Phys. Rev. \textbf{D78}, 064053
(2008); Saeed Mirshekari, Amir M. Abbassi, Int. J. Mod. Phys. \textbf{A24},
789 (2009); Saeed Mirshekari, Amir M. Abbassi, Mod. Phys. Lett. \textbf{A24},
747 (2009); E. C. Vagenas, JHEP \textbf{0307}, 046 (2003); E. C. Vagenas, Int.
J. Mod. Phys. \textbf{D14}, 573 (2005); Th. Grammenos, Mod. Phys. Lett.,
\textbf{A20}, 1741 (2005); T. Multamaki, A. Putaja, E. C. Vagenas and I.
Vilja, Class. Quant. Grav. \textbf{25}, 075017 (2008).

\bibitem {13}Gamal G. L. Nashed, Nuovo Cim. \textbf{117B}, 521 (2002); Gamal
G. L. Nashed, Phys. Rev. \textbf{D66}, 064015 (2002); Gamal G. L. Nashed, Int.
J. Mod. Phys. Lett. \textbf{A21}, 3181 (2006).

\bibitem {14}K. S. Virbhadra, Phys. Rev. \textbf{D60}, 104041 (1999). E. C.
Vagenas, Int. J. Mod. Phys\textit{.} \textbf{A18}, 5781 (2003); E. C. Vagenas,
\textbf{A18}, 5949 (2003); E. C. Vagenas, \textbf{D14}, 573 (2005); Th.
Grammenos, Mod. Phys. Lett\textit{.} \textbf{A20}, 1741 (2005).

\bibitem {15}R. Penrose, Proc. R. Soc. London, \textbf{A381}, 53 (1982).

\bibitem {16}K. P.Tod, Proc. R. Soc. London, \textbf{A388}, 457 (1983).

\bibitem {17}S. S. Xulu, Mod. Phys. Lett. \textbf{A15}, 1511 (2000); S. S.
Xulu, PhD Thesis, arXiv:hepth/0308070; S. S. Xulu, Astrophys. Space. Sci.
\textbf{283}, 23 (2003); I-Ching Yang, Ching-Tzung Yeh, Rue-Ron Hsu and
Chin-Rong Lee, Int. J. Mod. Phys. \textbf{D6}, 349 (1997); I-Ching Yang,
Wei-Fui Lin and Rue-Ron Hsu, Chin. J. Phys. \textbf{37}, 113 (1999); E.
C.Vagenas, Int. J. Mod. Phys. \textbf{A18}, 5949 (2003); E. C. Vagenas, Int.
J. Mod. Phys. \textbf{A18}, 5781 (2003); E. C. Vagenas, Mod. Phys. Lett.
\textbf{A19}, 213 (2004); E. C. Vagenas, Int. J. Mod. Phys. \textbf{D14}, 573
(2005); E. C. Vagenas, Mod. Phys. Lett. \textbf{A21}, 1947 (2006); Th.
Grammenos, Mod. Phys. Lett. \textbf{A20}, 1741 (2005); I. Radinschi, Mod.
Phys. Lett. \textbf{A15(35)}, 2171 (2000); I. Radinschi, Fizika \textbf{B9(4)}%
, 203 (2000); I. Radinschi, APH N.S., Heavy Ion Physics \textbf{12}, 47
(2000); I. Radinschi, Acta Physica Slovaca \textbf{50(4)}, 609 (2000); I.
Radinschi, Chin. J. Phys. \textbf{39(3)}, 231 (2001); I. Radinschi, Chin. J.
Phys. \textbf{39(5)}, 1 (2001); I-Ching Yang and I. Radinschi, Chin. J. Phys.
\textbf{41}, 326 (2003); I-Ching Yang, Chi-Long Lin and I. Radinschi, Int. J.
Theor. Phys. \textbf{48(1)}, 248 (2009); I-Ching Yang, Chi-Long Lin and I.
Radinschi, Int. J. Theor. Phys. \textbf{48(8)}, 2454 (2009).

\bibitem {18}I-Ching Yang, Chin. J. Phys. \textbf{45(5)}, 497 (2007); M.
Sharif and Tasnim Fatima, Int. J. Mod. Phys. \textbf{A20}, 4309 (2005); M.
Sharif, Nuovo Cim. \textbf{B19}, 463 (2004); M. Sharif, Int. J. Mod. Phys.
\textbf{D13}, 1019 (2004); M. Sharif and Tasnim Fatima, Nuovo Cim.
\textbf{B120}, 533 (2005); M. Sharif and Tasnim Fatima, Astrophys. Space Sci.
\textbf{302}, 217 (2006); M. Sharif, M. Azam, Int. J. Mod. Phys. \textbf{A22},
1935 (2007); Ragab M. Gad, Mod. Phys. Lett. \textbf{A19}, 1847 (2004); Ragab
M. Gad, Gen. Rel. Grav. \textbf{38}, 417 (2006); Ragab M. Gad, Astrophys.
Space Sci. \textbf{295}, 451 (2005); Ragab M. Gad, Astrophys. Space Sci.
\textbf{293}, 453 (2004); Ragab M. Gad, Astrophys. Space Sci. \textbf{295},
459 (2005); Ragab M. Gad, Astrophys. Space Sci. \textbf{302} 141 (2006); Ragab
M. Gad, Int. J. Theor.Phys. \textbf{46}, 3263 (2007); I. Radinschi and Th.
Grammenos, Int. J. Mod. Phys. \textbf{A21}, 2853 (2006); Th. Grammenos and I.
Radinschi, Int. J. Theor. Phys. \textbf{46}, 1055 (2007); I. Radinschi and Th.
Grammenos, Int. J. Theor. Phys. \textbf{47}, 1363 (2008); J. Matyjasek, Mod.
Phys. Lett. \textbf{A23(8)}, 591 (2008).

\bibitem {19}G. Lessner, Gen. Relativ. Gravit\textit{.} \textbf{28}, 527 (1996).

\bibitem {20}Chia-Chen Chang, J. M. Nester and Chiang-Mei Chen, Phys. Rev.
Lett. \textbf{83}, 1897 (1999).

\bibitem {21}S. Sur, S. Das and S. SenGupta, JHEP, 0510, 064 (2005) arXiv: hep-th/0508150

\bibitem {22}D. Garfinkle, G. T. Horowitz and A. Strominger, Phys. Rev.
\textbf{D43}, 3140 (1991), Phys. Rev. \textbf{D45}, 3888 (1992).

\bibitem {23}G. W. Gibbons, Nucl. Phys. \textbf{B207}, 337 (1982).

\bibitem {24}G. W. Gibbons and K. Maeda, Nucl. Phys. \textbf{B298}, 741 (1988).

\bibitem {25}K. C. K. Chan, J. H. Horne and R. B. Mann, Nucl. Phys.
\textbf{B447}, 441 (1995).

\bibitem {26}D. Gershon, Nucl. Phys. \textbf{B421}, 80 (1994).

\bibitem {27}A. Komar, Phys. Rev.\textbf{113}, 934 (1959).
\end{thebibliography}
\end{document}